# Instabilities at vicinal crystal surfaces – competition between the electromigration of the adatoms and the kinetic memory effect

Bogdan Ranguelov and Stoyan Stoyanov
"Rostislaw Kaischew" Institute of Physical Chemistry, Bulgarian Academy of Sciences,
"Acad. G. Bonchev" Str., Block 11, Sofia-1113, Bulgaria

Abstract
We studied the step dynamics during sublimation and growth when the adatoms on the crystal surface have a drift velocity $\frac{D_s F}{kT}$ where $D_s$ is the surface diffusion coefficient and $F$ is a force acting on the adatoms ( $F$ is related to the electric current heating the crystal ). In the limit of fast surface diffusion and slow kinetics of atom attachment-detachment at the steps we formulate a model free of the quasi-static approximation in the calculation of the adatom concentration on the terraces. The linear stability analysis of a step train results in an instability condition in the form $-(f\tau_s^/ /3\overline{\eta}) + (V/V_{cr}) - 1 > 0$ where $\tau_s^/$ is the dimensionless life-time of an adatom before desorption, $f$ and $\overline{\eta}$ are dimensionless electromigration force and the force of step repulsion whereas $V$ and $V_{cr}$ are the velocity of steps in the train and the critical velocity respectively. As seen instability is expected when either the velocity $V$ is larger than $V_{cr}$ ( this instability is related to the "kinetic memory effect" ), or $f\tau_s^/ /3\overline{\eta} > 1$, i.e when the electromigration force $f$ is negative (but strong enough to dominate over the term $V/V_{cr}$) which means step-down direction of the drift. In the latter case the initial stage of the step bunching process dramatically depends on the value of the ratio $-f/\overline{\eta}$. When $-f/\overline{\eta} \geq 1$ the instability starts with a formation of pairs (or small bunches) of steps, whereas at $-f/\overline{\eta} << 1$ the instability starts with a formation of relatively large bunches containing $N_{initial} = 2\pi/q_{max} = 2\pi/(-f/3\overline{\eta})^{1/2}$ steps. Numerical integration of the equations for the time evolution of the adatom concentrations and the equations of step motion reveals two different step bunching instabilities: 1) step density waves (small bunches which do not manifest any coarsening) induced by the kinetic memory effect and 2) step bunching with coarsening when the dynamics is dominated by the electromigration. The model developed in this paper also provides very instructive illustrations of the Popkov-Krug dynamical phase transition during sublimation and growth of a vicinal crystal surface.

## 1. Introduction

In a recent paper [1] we advanced a new model for the step dynamics during sublimation and growth of a vicinal crystal surface. This model contains the same physics as the classical Burton, Cabrera, Frank (BCF) theory [2, 3, 4] but the mathematical treatment deviates from the BCF procedure. In contrast to their assumption that the adatom concentrations $n_i$ on the terraces reach instantly their steady state for a given step configuration we analyzed the non-steady state problem [1]. This is relatively easy in the limiting case of fast surface diffusion and slow kinetics of atom attachment and detachment at the steps, since the fast diffusion provides for a constant value of the adatom concentration all over a given terrace. We derived equations for the time evolution of the adatom concentration on the terraces and, also, equations for the step motion. In this way we were able to treat accurately the case when the time to reach steady state concentration of adatoms on the terraces is compatible with the time for non-negligible change of the step configuration. In such a situation a new effect becomes important – the adatom concentration on a given terrace depends not only on the terrace size but on the "past of the terrace" as well. We call this a "kinetic memory effect". This effect provides a ground for a new type of instability of the regular step distribution. Step density compression waves appear at the vicinal surface as shown by both linear stability analysis and numerical integration of the equations for step motion [1]. Here our aim is to see how the "kinetic memory effect" competes with another (well known) destabilizing factor – the electromigration of the adatoms [5-11].

## 2. Basic equations

It is relatively easy to extend the treatment in Ref.[1] to account for the electromigration of the adatoms. The model is based on the following assumption - adatoms are subject of electromigration force (see Fig.1), so that they jump more frequently in the direction of this force than in the opposite direction [6, 12]. In addition one assumes the surface diffusion to be much faster than the atom attachment-detachment kinetics at the steps. Since the electromigration force pushes the adatoms towards one of the steps which separate the terrace from its neighbors, the concentration close to this step is higher than the average concentration over the whole terrace. Therefore more atoms attempt to attach to this step and become "crystal atoms". On the contrary, at the other step of the same terrace the concentration of adatoms is lower and less atoms attempt to join the crystal. This simple physics should be incorporated into the equations for the time evolution of the terrace widths. In a linear approximation the adatom concentration on a given terrace is $n_i(\xi,t) = n_i(t)\left(1 + \frac{F\xi}{kT}\right)$ with $\xi = 0$ being in the middle of the terrace. In this approximation $n_i(t)$ is equal to the average value of the adatom concentration on the terrace.

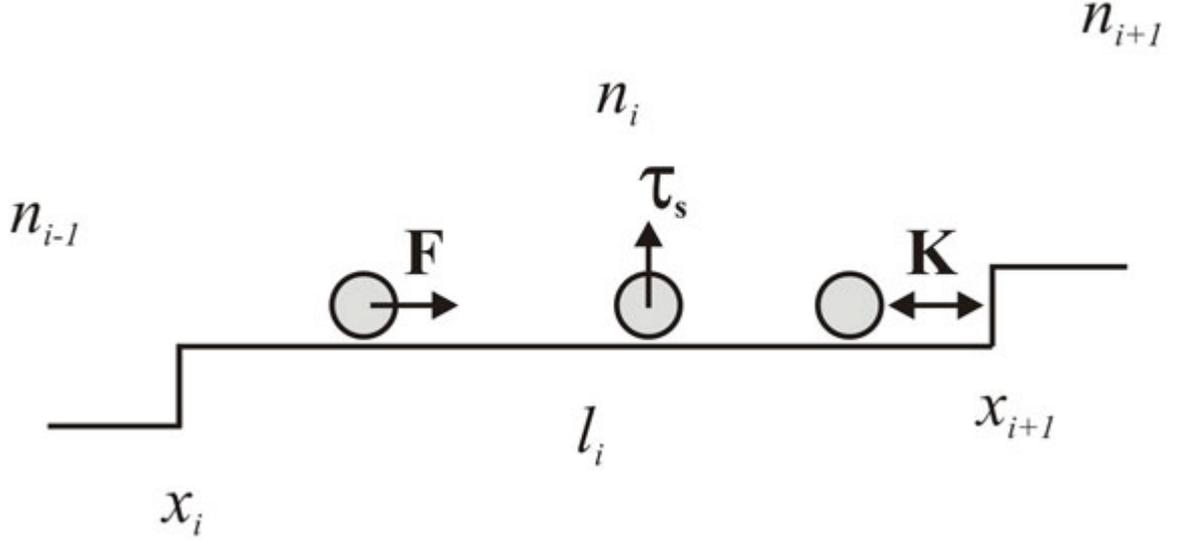

FIG. 1. Schematic view of a vicinal surface with straight steps. The adatoms have a drift velocity induced by an electric force $F$. The width of the $i-th$ terrace is $l_i = x_{i+1} - x_i$ and the concentration of adatoms on it is $n_i$. The desorption flux from the terrace is $n_i/\tau_s$ and the exchange between the crystal phase and the dilute layer of adatoms is characterized by the step kinetic coefficient $K$.

The space non-uniformity of the adatom concentration has an impact on the rates of step motion. Really, now we have

$$v_i = -K\Omega\left[n_i\left(1 - \frac{Fl_i}{2kT}\right) - n_s^e(i)\right] - K\Omega\left[n_{i-1}\left(1 + \frac{Fl_{i-1}}{2kT}\right) - n_s^e(i)\right] \quad (1)$$

where $K$ is the step kinetic coefficient (see Fig.1) and $\Omega$ is the area of one atomic site at the crystal surface. On the basis of eq.(1) one can write expression for the time derivatives of the width of the terraces

$$\frac{dl_i}{dt} = -\Omega\left[\begin{array}{l} K\left(n_{i+1}\left(1 - \frac{Fl_{i+1}}{2kT}\right) - n_s^e(i+1)\right) + K\left(n_i\left(1 + \frac{Fl_i}{2kT}\right) - n_s^e(i+1)\right) \\ - K\left(n_{i-1}\left(1 + \frac{Fl_{i-1}}{2kT}\right) - n_s^e(i)\right) - K\left(n_i\left(1 - \frac{Fl_i}{2kT}\right) - n_s^e(i)\right) \end{array}\right] \quad (2)$$

where $n_i$ are functions of time.

As known the equilibrium concentration of adatoms in the vicinity of the steps has the constant value $n_s^e$ only when $l_i = l_{i-1} = l$ with $l$ being the average terrace width determined by the miscut angle of the vicinal surface. In the general case one has [13, 14]

$$n_s^e(i) = n_s^e\left[1 + \tilde{A}\left(\frac{1}{l_{i-1}^3} - \frac{1}{l_i^3}\right)\right] \quad (3)$$

where $\tilde{A} = \dfrac{2\Omega A}{kT}$ and A is the strength of the entropic and stress-mediated repulsion between the steps.

Introducing dimensionless variables $\tau = \dfrac{Kt}{l}$, $\eta_i = \dfrac{l_i}{l}$, $c_i = \dfrac{n_i}{n_s^e}$, $\bar{\eta} = \dfrac{\tilde{A}}{l^3}$, we obtain for the dimensionless width $\eta_i(\tau)$ of the $i-th$ terrace (details on the derivation of these equations can be found in [1])

$$\dfrac{d\eta_i}{d\tau} = -n_s^e \Omega \left\{ c_{i+1} - c_{i-1} - f(\eta_{i+1}c_{i+1} - 2\eta_i c_i + \eta_{i-1}c_{i-1}) + 2\bar{\eta}\left(\dfrac{1}{\eta_{i+1}^3} - \dfrac{2}{\eta_i^3} + \dfrac{1}{\eta_{i-1}^3}\right) \right\} \quad (4)$$

where $f = \dfrac{Fl}{2kT}$. As for the equations for the time evolution of the dimensionless concentrations of adatoms, they are not affected by the electromigration force, i.e. they read [1]

$$\dfrac{dc_i}{d\tau} = -\dfrac{c_i}{\tau_s'} - \dfrac{2}{\eta_i}c_i + \dfrac{2}{\eta_i} + \dfrac{1}{\eta_i}\bar{\eta}\left(\dfrac{1}{\eta_{i-1}^3} - \dfrac{1}{\eta_{i+1}^3}\right) \quad (5)$$

where $\tau_s' = \dfrac{\tau_s K}{l}$ and $\tau_s$ is the average time for desorption of an atom in a state of mobile adsorption on the crystal surface (see Fig.1). The equations (5) for the dimensionless concentration have a clear physical meaning – the adatom concentration decreases because of the desorption (the first term in the right hand side) and attachment of adatoms to the steps (the second term). The concentration $c_i$ of adatoms increases because of the detachment of atoms from the steps (the next terms in the right hand side of Eq.(5)). One should keep in mind however, that now $c_i(\tau)$ is the average value of the dimensionless adatom concentration on the $i-th$ terrace.

In a presence of non-zero deposition rate $R$ an additional term appears in the right hand side of eq. (5) - this term reads $\dfrac{c_{st}}{\tau_s'}$ where $c_{st} = \dfrac{R\tau_s}{n_s^e}$. Obviously, when the deposition rate has its equilibrium value $R_e$ one obtains $c_{st} = \dfrac{R_e \tau_s}{n_s^e} = 1$. This means that $c_{st} < 1$ corresponds to undersaturation and $c_{st} > 1$ corresponds to supersaturation at the crystal surface. It is instructive to write down the equations for the dimensionless concentrations on the terraces in a presence of deposition rate

$$\dfrac{dc_i}{d\tau} = \dfrac{c_{st}}{\tau_s'} - \dfrac{c_i}{\tau_s'} - \dfrac{2}{\eta_i}c_i + \dfrac{2}{\eta_i} + \dfrac{1}{\eta_i}\bar{\eta}\left(\dfrac{1}{\eta_{i-1}^3} - \dfrac{1}{\eta_{i+1}^3}\right) \quad (6)$$

3. Linear stability analysis.

The simplest solution of the equations (4) and (5) is $\eta_i = 1$ and $c_i = c_0 = \dfrac{1}{1+1/2\tau'_s}$ which is an equidistant step distribution and the corresponding constant concentration of adatoms. The stability of this solution with respect to small fluctuations of the terrace size and adatom concentration is an interesting problem. To solve it we follow the routine procedure and consider small deviations from the equidistant step train and constant adatom concentrations, i.e., $\eta_i = 1 + \Delta\eta_i(\tau)$ and $c_i = c_0 + \Delta c_i(\tau)$. Substituting these expressions into Eqs. (4) and (5), making use of series expansion and keeping the linear terms only we get

$$\frac{d\Delta\eta_i}{d\tau} = -n_s^e \Omega[\Delta c_{i+1} - \Delta c_{i-1} - (fc_0 + 6\overline{\eta})(\Delta\eta_{i+1} - 2\Delta\eta_i + \Delta\eta_{i-1}) - f(\Delta c_{i+1} - 2\Delta c_i + \Delta c_{i-1})] \quad (7)$$

$$\frac{d\Delta c_i}{d\tau} = -\frac{\Delta c_i}{\tau'_s} - 2\Delta c_i + 2(c_0 - 1)\Delta\eta_i + 3\overline{\eta}(\Delta\eta_{i+1} - \Delta\eta_{i-1}) \quad (8)$$

Following the routine we look for a solution of the type $\Delta\eta_j = e^{ijq}\eta_q(\tau)$ and $\Delta c_j = e^{ijq+i\phi}c_q(\tau)$, where $q$ is a wave number and we already use $i$ to denote the imaginary unit and $j$ to denote the sequence number of the terrace. In addition we allow for a phase shift $\phi$ of the wave describing the adatom concentrations $\Delta c_j$ with respect to the wave describing the terrace widths $\Delta\eta_j$. In this way we arrive to a set of two differential equations for the time evolution of the amplitudes of the fluctuations in the terrace width distribution and adatom concentrations

$$\frac{d\eta_q}{d\tau} = a_{11}\eta_q(\tau) + a_{12}c_q(\tau) \quad (9)$$

$$\frac{dc_q}{d\tau} = a_{21}\eta_q(\tau) + a_{22}c_q(\tau)$$

where the coefficients are given by the following expressions
$$\begin{aligned}
a_{11} &= -2n_s^e\Omega(6\overline{\eta} + fc_0)(1-\cos q) \\
a_{12} &= -2e^{i\phi}n_s^e\Omega[i\sin q + f(1-\cos q)] \\
a_{21} &= 2e^{-i\phi}[-(1-c_0) + 3i\overline{\eta}\sin q] \\
a_{22} &= -(2+1/\tau'_s)
\end{aligned} \quad (10)$$

This set of two linear differential equations has a solution of the type $e^{s\tau}$ where $s$ satisfies
$$(a_{11} - s)(a_{22} - s) - a_{12}a_{21} = 0 \quad (11)$$
For the real part of $s$ one obtains (technical details on the derivation of a similar expression when $f = 0$ are given in [1])

$$s_r = 2n_s^e \Omega \left[ \begin{array}{c} -(6\overline{\eta} + fc_0)(1-\cos q) + 2\dfrac{3\overline{\eta}\sin^2 q + f(1-c_0)(1-\cos q)}{(a_{11}-a_{22})} + \\ 8\dfrac{(n_s^e \Omega)\sin^2 q[-(1-c_0)+3\overline{\eta}f(1-\cos q)]^2}{(a_{11}-a_{22})^3} \end{array} \right] \qquad (12)$$

Restricting the treatment to small wave numbers we use series expansions $\sin q \approx q - \dfrac{1}{3!}q^3$ and $\cos q \approx 1 - \dfrac{1}{2!}q^2 + \dfrac{1}{4!}q^4$ and substitute them into the last expression. Thus for the limit $\tau_s' \gg 1$ we obtain

$$s_r = B_2 q^2 - B_4 q^4 \qquad (13)$$

where

$$B_2 = \dfrac{3n_s^e \Omega \overline{\eta}}{\tau_s'}\left[-1-\dfrac{f\tau_s'}{3\overline{\eta}}+\dfrac{n_s^e \Omega}{6\overline{\eta}\tau_s'}\right] \qquad (14)$$

$$B_4 = \dfrac{3n_s^e \Omega \overline{\eta}}{\tau_s'}\left[\dfrac{\tau_s'}{2}-\dfrac{f\tau_s'}{36\overline{\eta}}+\dfrac{n_s^e \Omega}{18\tau_s'\overline{\eta}}\right] \qquad (15)$$

As seen the equations (13) - (15) can be reduced to the already known expressions (see Ref.[1]) when the electromigration force $F$ is zero and, therefore, $f = \dfrac{Fl}{2kT} = 0$. In Ref.[1] we used the notation $V = n_s^e \Omega l / \tau_s$ for the velocity of steps in a perfectly regular train and $V_{cr} = 6K\overline{\eta}$ for the critical velocity.

Now we are in a position to discuss the impact of the electromigration on the step dynamics. To have instability of the step train (at small wave numbers, at least) we need a positive value of the constant $B_2$ (see eq.(13)). This could be achieved in two ways – either $\dfrac{n_s^e \Omega}{6\overline{\eta}\tau_s'} = \dfrac{V}{V_{cr}}$ is large enough to insure a positive sign of the expression in the brackets of Eq.(14), or $f < 0$ and the absolute value of $\dfrac{f\tau_s'}{3\overline{\eta}}$ dominates the term $\dfrac{n_s^e \Omega}{6\overline{\eta}\tau_s'}$ and is large enough to compensate -1. In the former case we expect step density compression waves to propagate at the vicinal surface as shown in [1]. In the latter case the step dynamics is expected to be dominated by the adatom electromigration effect. As for the value of $B_4$ it is definitely positive at $f < 0$ (see eq.(15)).

It is interesting to note that the eq. (14) defines a critical value of the electromigration force. Really, instability induced by electromigration takes place only when $\frac{f\tau_s'}{3\bar{\eta}} > 1$ which is equivalent to $F > F_c = \frac{12\Omega A}{K\tau_s l^3}$. This expression for $F_c$ has been published by Uwaha et al [15].

In a presence of deposition rate the linear stability criterion is slightly modified in a sense that the expression for $B_2$ now contains the quantity $c_{st} = \frac{R\tau_s}{n_s^e}$. Simple considerations result

$$B_2 = \frac{3n_s^e \Omega \bar{\eta}}{\tau_s'} \left[ -1 - \frac{f\tau_s'}{3\bar{\eta}}\left(1 + \frac{c_{st}}{\tau_s'}\right) + \frac{n_s^e \Omega}{6\bar{\eta}\tau_s'}(1-c_{st})^2 \right] \quad (16)$$

Since the modifying factors $\left(1 + \frac{c_{st}}{\tau_s'}\right)$ and $(1-c_{st})^2$ are always positive the deposition rate does not change qualitatively the instability criterion. Step bunching with coarsening occurs at $f < 0$, i.e. step down direction of the electric force when the step dynamics is dominated by the atom electromigration. A formation of small bunches which do not show coarsening takes place in both growth ($c_{st} > 1$) and sublimation ($c_{st} < 1$) when the dominating factor is the "kinetic memory effect". The deposition rate is expected to have an impact only on the transition from one type of instability to the other one.

Finally for the wave number of the most unstable mode (the maximum value of $s_r$) during electromigration dominated sublimation one obtains the approximate expression

$$q_{max} = \left[\frac{B_2}{2B_4}\right]^{1/2} \approx \left[\frac{-\frac{f}{3\bar{\eta}}}{1 - \frac{f}{18\bar{\eta}}}\right]^{1/2} \quad (17)$$

This expression has a simple limit - at $\left|\frac{f}{18\bar{\eta}}\right| \ll 1$ the most unstable mode has a wave number

$$q_{max} = \left[-\frac{f}{3\bar{\eta}}\right]^{1/2} \quad (18)$$

which is also small, i.e. the instability starts with fluctuations having a large wave length. The last expression provides a ground to conclude that the wavelength of the most unstable mode depends on the average terrace width $l$ of the vicinal surface. Really, having in mind the definitions of $f$ and $\bar{\eta}$ we find $q_{max} \sim l^2$ and therefore the corresponding wavelength of the most unstable mode

$$\lambda_{max} = j_{max} l = \frac{2\pi}{q_{max}} l \sim \frac{1}{l} \quad (19)$$

is inversely proportional to the average terrace width.

According to Eq.(17) the increase of $\left|\dfrac{f}{18\bar{\eta}}\right|$ leads to an increase of the wave number $q_{max}$ and the step bunching instability starts with a formation of very small bunches or pairs of steps which later coalesce to form larger bunches.

4. Long time behavior of vicinal surfaces

We study the non-linear dynamics of the steps by numerical integration of the equations (4) and (5). Our aim is to look for a confirmation of the results obtained by the linear stability analysis. On the other hand, it is interesting to explore the capability of this model (free of the quasi-static approximation) to manifest the dynamic phase transition predicted by Popkov and Krug [16].

The set of equations (4) and (5) involves 2N variables because we use circling boundary conditions $\eta_{N+1} = \eta_1$ and $c_{N+1} = c_1$. To obtain instructive picture of the sublimation dynamics we calculate the step trajectories in a frame moving with the average velocity of the step train (containing N steps). The calculated trajectories clearly show (Fig.2) the propagation of step density waves (due to "kinetic memory effect) and a formation of bunches of steps (Fig.3) due to the electromigration of the adatoms. Some comments on these two instabilities will be made on the basis of eq.(14). Substituting the values of the model parameters $\tau_s^/ = 100$, $\bar{\eta} = 10^{-5}$, $n_s^e \Omega = 0.1$ and $f = -10^{-4}$ (which we used to obtain the trajectories shown in Fig.2) into eq.(14) we see that $-\dfrac{f \tau_s^/}{3\bar{\eta}} = 333.3$ and $\dfrac{n_s^e \Omega}{6\bar{\eta}\tau_s^/} = 16.6$.

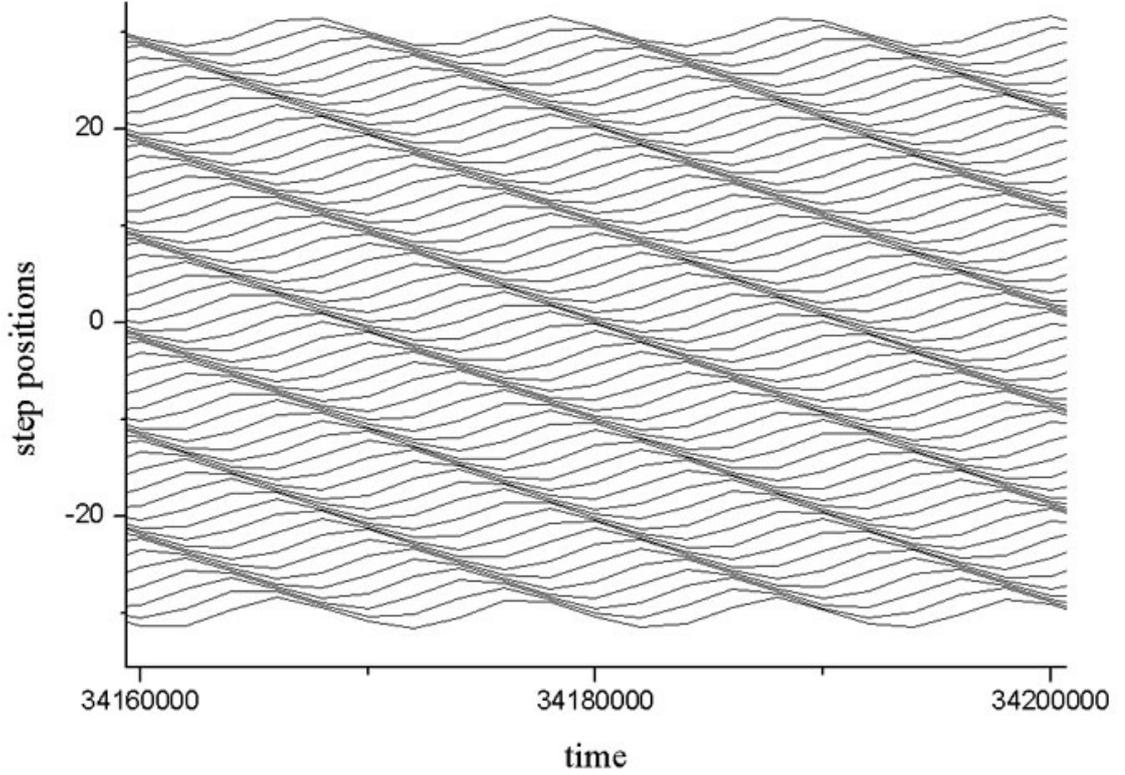

FIG.2. Step trajectories in a frame moving with a velocity averaged over all steps in the system (containing 60 steps). Compression waves propagate when the model parameters are $\tau_s^/ = 100$, $\bar{\eta} = 10^{-5}$, $n_s^e \Omega = 0.1$ and $f = -10^{-4}$.

Although the term containing the electromigration force is 20 times larger than the term related to "kinetic memory effect" the non-linear dynamics manifests step density waves, i.e., the dynamics is dominated by the "kinetic memory effect". Therefore, the linear stability analysis does not provide reliable conclusion about the dominating instability – the above estimations of the terms into eq.(14) predicts the electromigration induced bunching of steps to develop faster than the step density waves, but this is not the case.

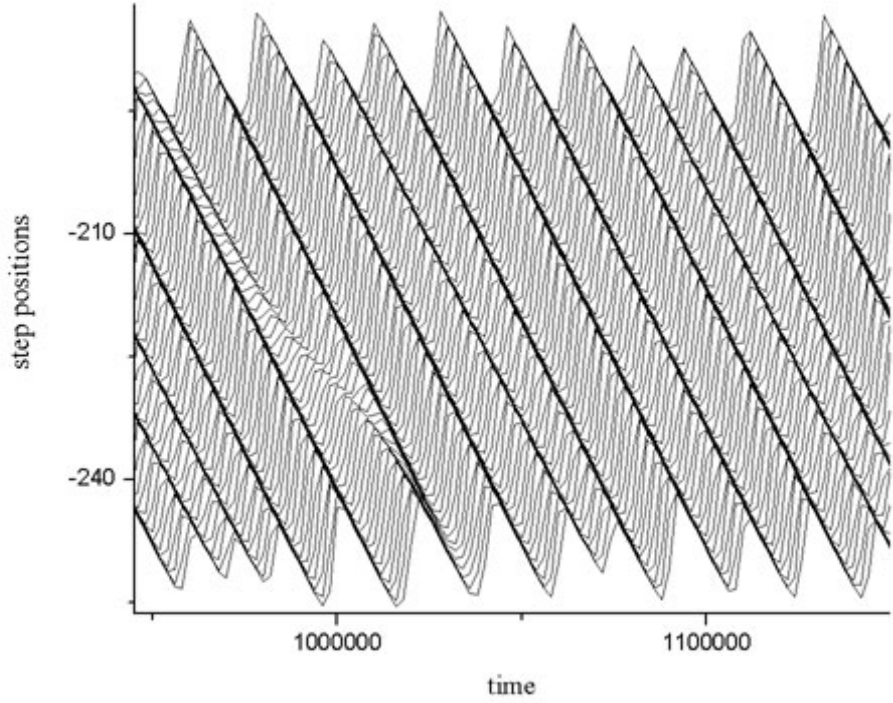

FIG.3. Step trajectories manifest coarsening - step bunches grow in size with the sublimation time. The model parameters are $\tau'_s = 100$, $\bar{\eta} = 10^{-5}$, $n^e_s \Omega = 0.1$ and $f = -10^3$.

To study the transition from step density waves instability to step bunching (with coarsening) instability we gradually increased the absolute value of the parameter $f$ and integrated the equations (4) and (5) at $f = -2 \times 10^{-4}$ and $f = -3 \times 10^{-4}$. The results manifest a formation of step density waves which display very slow coarsening (at $f = -3 \times 10^{-4}$ coarsening is a bit faster). Further increase of the absolute value of the parameter $f$ (see Fig.3) leads to coarsening with substantial rate.

It is interesting to check the results of the linear stability analysis for the most unstable mode. The Eq.(17)-(19) and the discussion after them predicts the bunching process to start either with a formation of pairs or with a formation of much larger bunches depending on the value of the parameter $\left|\dfrac{f}{18\bar{\eta}}\right|$. A good illustration of long wavelength instability is provided by Fig.4 which shows the step trajectories when $\tau'_s = 2000$, $\bar{\eta} = 10^{-3}$, $n^e_s \Omega = 0.1$ and $f = -2 \times 10^{-3}$. Substituting these values into the ratio $\left|\dfrac{f}{18\bar{\eta}}\right|$ one obtains $\dfrac{1}{9}$ which is much smaller than unity so that the wave number of the most unstable mode is given by eq.(18). Thus one obtains $q_{max} = [-f/3\bar{\eta}]^{1/2} = \sqrt{\dfrac{2}{3}} = 0.816$. The corresponding wavelength measured in number of terraces is $\lambda_{max} \sim j_{max} = \frac{2\pi}{q_{max}} \approx 7.7$ in good agreement with the results shown in Fig.4.

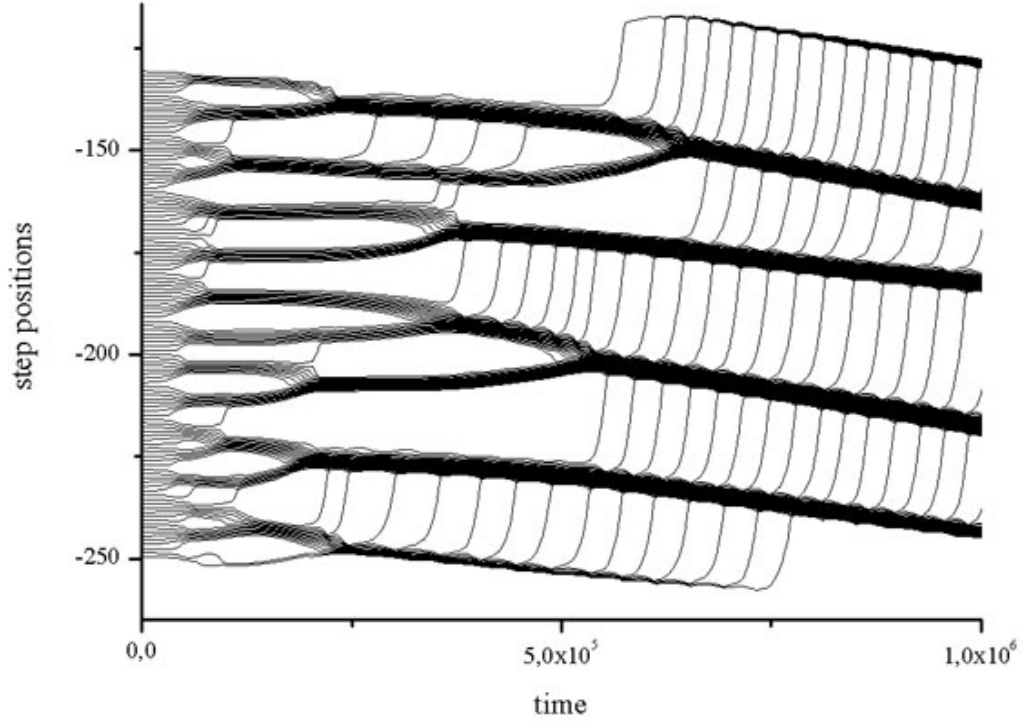

FIG.4. The step trajectories obtained at $n_s^e \Omega = 0.1; f = -2.0 \times 10^{-3}; \bar{\eta} = 10^{-3}; \tau_s' = 2000$ manifest three interesting features of the step dynamics – 1) long wavelength instability; 2) existence of dynamic regime with all steps involved in bunches [16]; 3) existence of critical size above which the bunch starts to emit single steps [16].

Finally, it is interesting to explore the capability of this model to manifest dynamic phase transition predicted by Popkov and Krug [16] on the basis of the quasi-static approximation. The dynamic phase transition is, in fact, a transition between two regimes of step dynamics which are qualitatively different. In the most clear situation one of the regimes is characterized by an absence of any single steps at the crystal surface, i.e. all steps are in bunches (such a situation is shown in Fig.4 for relatively short time of evaporation). The second regime can be described as a co-existence of bunches and single steps crossing the terraces (essential requirement is to have several steps crossing simultaneously each terrace (see Fig.3)). Popkov and Krug found the transition to take place at the critical value $b=1$ of the asymmetry parameter $b$ defined by the expression [16]

$$\frac{dx_i}{dt} = \frac{1-b}{2}(x_{i+1} - x_i) + \frac{1+b}{2}(x_i - x_{i-1}) + U \qquad (20)$$

where $x_i$ is the position of the $i-th$ step, $U$ accounts for the step-step repulsion and the time scale is normalized to the sublimation rate. To express $b$ through the parameters of our model we rewrite the equation (1) to introduce dimensionless parameters and make use of the steady state solution $c_i = \frac{1}{1+\eta_i/2\tau_s^/}$ of Eq.(5). Thus for the limit $\tau_s^/ \gg 1$ one obtains the following expression for the step rate during sublimation (the deposition rate is zero)

$$\frac{dx_i}{dt} = -K\Omega n_s^e \{[c_i(1-f\eta_i)-1]+[c_{i-1}(1+f\eta_{i-1})-1]\} = $$
$$K\Omega n_s^e \left\{\frac{\eta_i}{2\tau_s^/}(1+2f\tau_s^/)+\frac{\eta_{i-1}}{2\tau_s^/}(1-2f\tau_s^/)\right\} \qquad (21)$$

After appropriate normalization of the time scale the last equation can be transformed into eq.(20) with $b = -2f\tau_s^/$.

Our numerical integration of eqs.(4) and (5) reproduces the results of Popkov and Krug [16]. Figure 4 manifests the existence of a regime of step dynamics where all steps are in the bunches (there are no steps crossing the terraces) at $b = -2f\tau_s^/ = 8$. Fig.4 obtained at values $n_s^e\Omega = 0.1; f = -2.0\times10^{-3}; \bar{\eta} = 10^{-3}; \tau_s' = 2000$ manifests another interesting feature of the step dynamics, predicted by Popkov and Krug [16] – when the bunch exceeds some critical size it starts to emit single steps. It is essential to say that Popkov and Krug used a somewhat softer definition of the "bunched phase" – steps crossing the terraces are not completely excluded. In their original paper [16] the requirement is to have no more than one step crossing the terrace at a time. That is why the larger value $(b = 8)$ of the asymmetry parameter used to obtain Fig.4 is not in contradiction with the value $(b = 1)$ derived in [16].

5. Instabilities during growth.
We studied the non-linear dynamics of steps in a presence of deposition rate by integrating the equations (4) and (6) for $\tau_s^/ = 10^3$, $\bar{\eta} = 10^{-3}$, $n_s^e\Omega = 0.01$ and $f = -5\times10^{-2}$ $c_{st} = 10^3$. In this way we obtained bunching with coarsening as seen in Fig.5. This figure contains very interesting results. Step trajectories show initial formation of pairs (short wavelength instability) which later coalesce to form larger bunches (hierarchical bunching). Single steps detach from the bunches and quickly cross the terraces to join the bunch ahead. This familiar picture of co-existing bunches and single steps is suddenly transformed into a single bunch containing all 60 steps in the system studied by numerical integration of the equations of step dynamics. This sharp transformation is a very instructive manifestation of the Popkov-Krug dynamical phase transition, although it is "reversed": from bunches exchanging steps to bunch(es) that do not exchange steps.

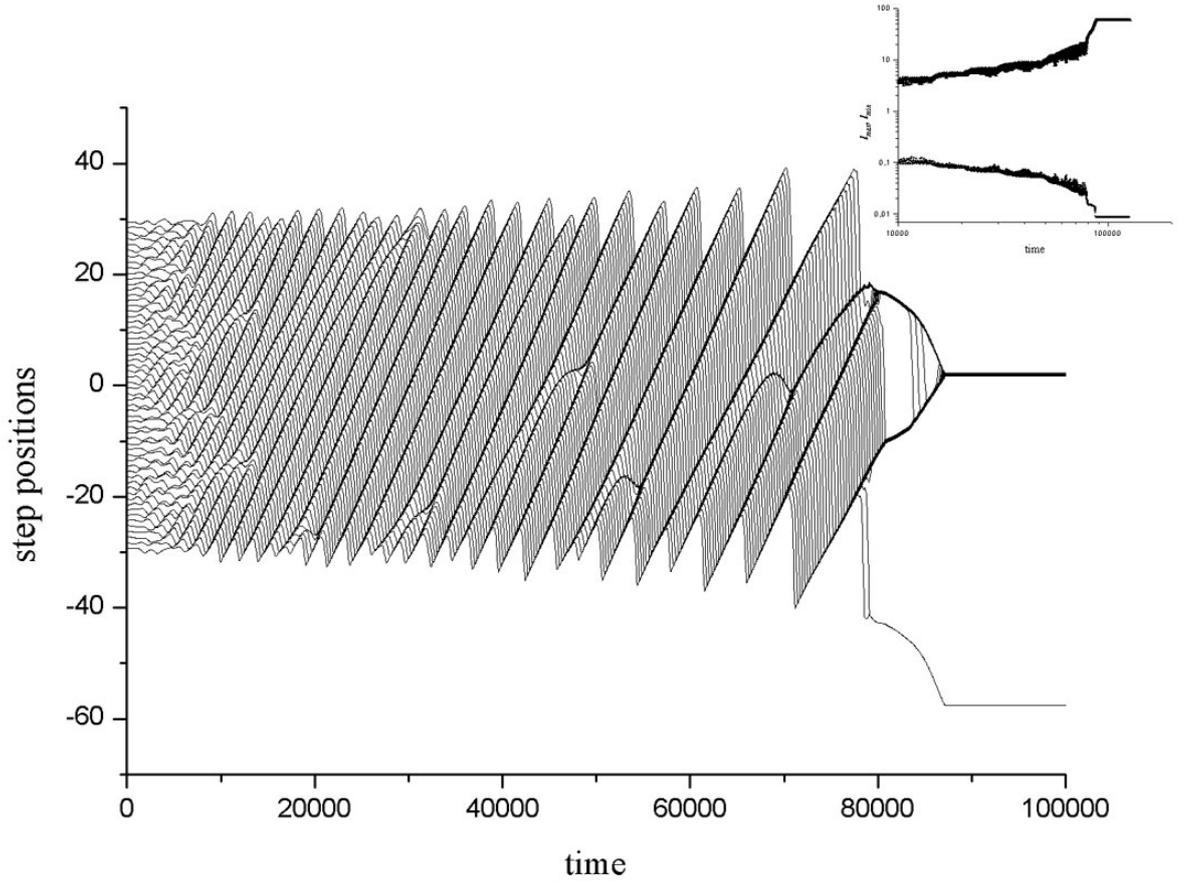

FIG. 5. Step trajectories during crystal growth (the model parameters are $\tau'_s = 100$, $\bar{\eta} = 10^{-3}$, $n_s^e \Omega = 0.01$, $c_{st} = 50$ and $f = -0.04$) clearly manifest a dynamic phase transition [16]. This transition is accompanied by a remarkable compression of the bunches (see the inset giving the maximum and the minimum interstep distance in the bunch).

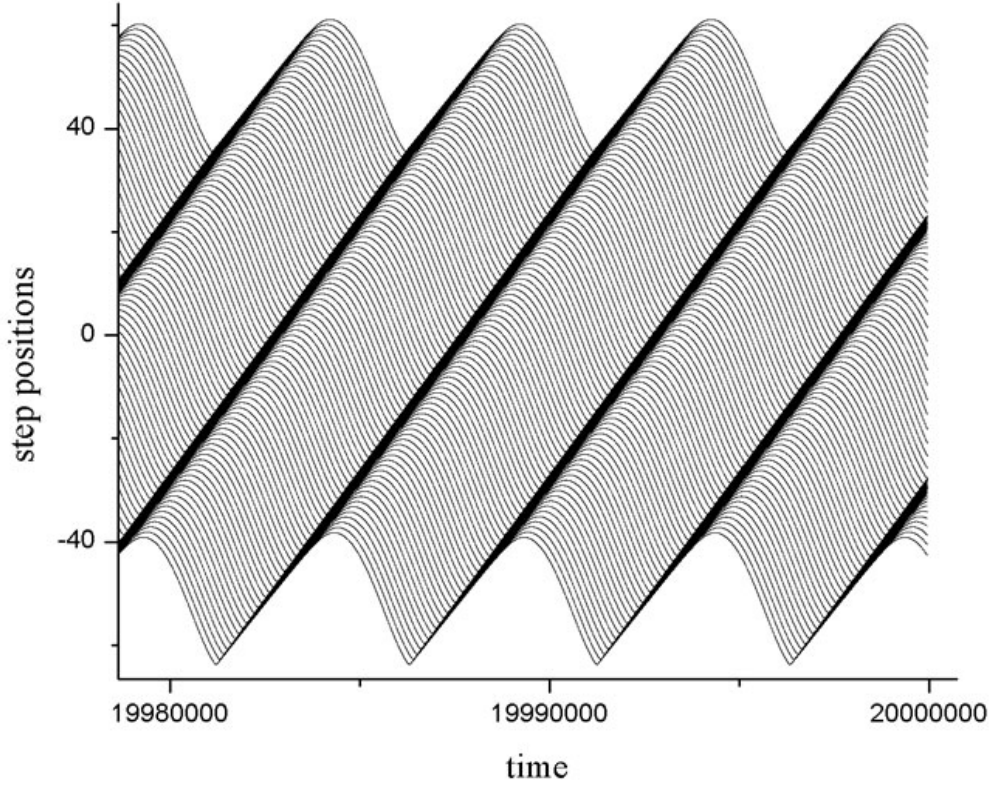

FIG. 6. Step trajectories in absence of electromigration $f = 0$. The model parameters have the values $\tau_s^i = 10^4, c_{st} = 10^4, \bar{\eta} = 10^{-3}, n_s^e \Omega = 0.01$. Relatively large step density waves (bunching without coarsening) are clearly seen.

Finally, it is interesting to study the step dynamics at a vicinal surface during crystal growth ( deposition rate larger than the equilibrium one) in absence of electromigration of adatoms. As seen in Fig.6, relatively large step density waves (bunching without coarsening) propagate at the vicinal surface.

6. Conclusions

Going beyond the quasi-static approximation in the treatment of step dynamics we see the vicinal surface to manifest two different step bunching instabilities. Formation of small bunches of constant size (absence of any coarsening) occurs when the electromigration is relatively weak and the step dynamics is dominated by the "kinetic memory effect" (see Fig.2 and 6). This is, in fact, the instability (formation of step density waves) reported in our previous paper [1]. Another type of instability (step bunches manifest coarsening so that the average bunch size increases with the sublimation time) occurs when the electromigration of adatoms is the dominating factor and the "kinetic memory effect" has a secondary role in the step dynamics. It is interesting to note that these two factors play identical roles in the linear stability theory where the instability condition reads $-(f\tau_s^/ / 3\overline{\eta}) + (V/V_{cr}) - 1 > 0$ (see Eqs.(13) and (14)). As far as the non-linear dynamics of steps is concerned these two factors have rather different impact. The asymmetry factor $b = -2f\tau_s^/$ leads to the formation of large bunches which manifest clear coarsening, whereas the deviation from the equilibrium ($V >> V_{cr}$) produces a regular array of small bunches of constant size (there is no coarsening at all).

The step bunching due to the asymmetry factor $b = -2f\tau_s^/$ is similar to the already studied [17] instability induced by electromigration in the limit of slow step kinetics and fast surface diffusion (the model in [17] makes use of the quasi-static approximation). Analyzing this instability we reproduce already published (by other authors) results for this limit. For example our Eq.(18) coincides with the Eqs.(33) and (60) in [18] and Eq.(4) in [19]. These results of the linear stability analysis for the most unstable mode are supported by Fig.3 and Fig.4 showing the step trajectories obtained by numerical integration of the equations (4) and (5).

Finally, our non-steady state treatment of the step dynamics at the crystal surface manifests the dynamic phase transition predicted by Popkov and Krug for the model based on the quasi-static approximation [16].